\documentclass[sigconf]{acmart}
\AtBeginDocument{%
  \providecommand\BibTeX{{%
    \normalfont B\kern-0.5em{\scshape i\kern-0.25em b}\kern-0.8em\TeX}}}

\copyrightyear{2023}
\acmYear{2023}
\setcopyright{rightsretained}
\acmConference[CHI EA '23]{Extended Abstracts of the 2023 CHI Conference on Human Factors in Computing Systems}{April 23--28, 2023}{Hamburg, Germany}
\acmBooktitle{Extended Abstracts of the 2023 CHI Conference on Human Factors in Computing Systems (CHI EA '23), April 23--28, 2023, Hamburg, Germany}\acmDOI{10.1145/3544549.3585741}
\acmISBN{978-1-4503-9422-2/23/04}

\begin{document}

\title{3D Printing and Design in Isolation: A Case from a Simulated Lunar Mission}


\author{Wiktor Stawski}
\affiliation{%
 \institution{XR Lab, Polish-Japanese Academy of Information Technology}
   \city{Warsaw}
 \country{Poland}}

\author{Kinga Skorupska}
\affiliation{%
 \institution{XR Lab, Polish-Japanese Academy of Information Technology}
   \city{Warsaw}
 \country{Poland}}

\author{Wiesław Kopeć}
\affiliation{%
 \institution{XR Lab, Polish-Japanese Academy of Information Technology}
   \city{Warsaw}
 \country{Poland}}

\renewcommand{\shortauthors}{Stawski, et al.}

\begin{abstract}

Despite the decades-long history of 3D printing, it is not used to its full potential. Yet 3D printing holds promise for isolated communities, aiming for self-sufficiency. In this experiential study conducted in an analog space habitat we evaluated challenges and opportunities of using 3D printing. Our study revealed barriers such as: 1) setting up and maintaining the 3D printing equipment while minding different kinds of pollution, that is air, temperature and sound, 2) design skill and familiarity with specialized software as well as materials and 3) the awareness of what can be achieved to meet community needs. We observed that in-community experience and know-how are reliable sources of 3D print ideas, that improve quality of life of community members if they are encouraged and supported by participatory design. Co-design of 3D prints in small, specialized communities is a promising area of study, that can bring new applications of 3D print technology. 

\end{abstract}

\ccsdesc[500]{Human-centered computing~Participatory design}
\ccsdesc[300]{Social and professional topics~Professional topics}
\ccsdesc[300]{Applied computing~Collaborative learning}
\keywords{3D printing, experiential study, 3D print lab setup, barriers to design, co-design, participatory design, isolated communities, analog astronauts}

\begin{teaserfigure}
  \includegraphics[width=\textwidth]{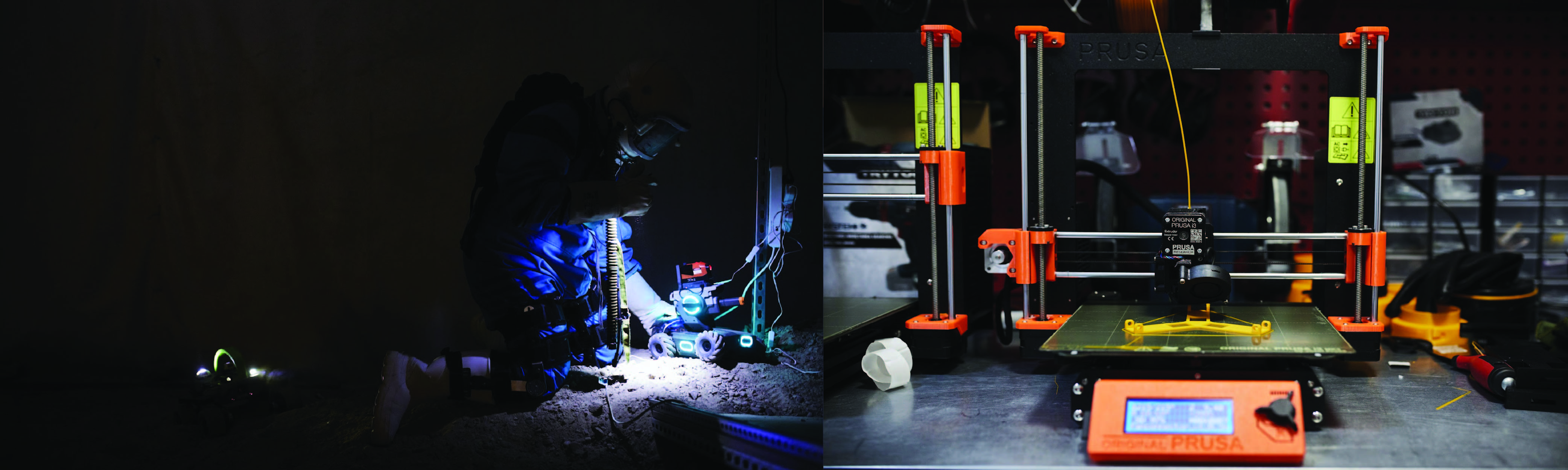}
  \caption{Left: Simulated Moon Environment in the Lunar Habitat, Right: Simulated Lunar Mission's 3D printing lab, where we conducted our experiential study. Photos by Susanne Baumann.}
  \Description{An analog astronaut, wearing a blue jumpsuit repurposed from a MIG pilot's outfit, is performing an EVA procedure with a robotic/plane on the left side. On the right side, a PRUSA 3D printer is producing a component for a project during the Alpha-XR mission.}
  \label{fig:teaser}
\end{teaserfigure}

\maketitle

\section{Rationale}

3D printing, despite its strangely prototypical status for a 30+-year-old technology, is widely used in commercial, private or hobbyist contexts. It is becoming increasingly popular among experienced users as well as enthusiastic novices. Invented by Chuck Hull, it is revolutionizing not only manufacturing but also industries such as interior design and architecture. Some say that 3D printing technology will even revolutionize the space industry. And it already is, as exemplified by the company RocketLab\footnote{More on the company can be found here: https://www.rocketlabusa.com/}, which is printing suborbital rocket engines and full shells in metal with the ambition of creating the first 3D printed orbital rocket. The value of 3D printing is especially evident when it comes to producing necessary products or replacement parts on site, in a low-cost way \cite{smallfarmsOpensource2015}, in remote locations where speedy transport is difficult (i.e. arctic stations), or plainly impossible (i.e. space exploration) \cite{https://doi.org/10.1002/ad.1840}, with the use of local materials (i.e. regolith \cite{MEURISSE2018800regolith}), especially in communities aiming for self-sufficiency \cite{KADING2015317}, even including those without access to electricity, thanks to solar-powered printers \cite{solar3Dprinters2016}. It also allows products to be tailored to individual needs and conditions, and to adjust designs on-site, which is particularly important for isolated communities whose experiences and needs are unique. The introduction of 3D printing to isolated communities allows for the delivery of various needed products and services (e.g. 3D printed food \cite{3Dfoodprintin2022}) without the need to physically deliver them to the site. \textbf{Hence, in this case study we evaluated how such isolated groups without much training may make use of 3D printing to begin to address their immediate needs, wants and desires and what are the challenges of using 3D printers in such conditions. Based on this experience we present lessons learned, including guidelines and recommendations for preparing working space, environments and knowledge-transfer opportunities for such groups to make better use of 3D printing.}

\section{Methods}

 \begin{figure*}[htb!]
 \centering
\includegraphics[width=\linewidth]{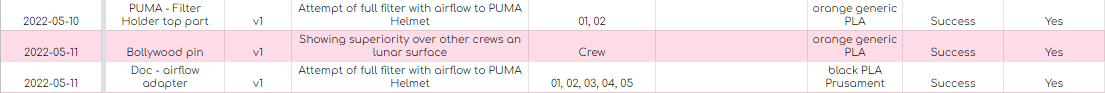}
 \caption{Excerpt from the 3D Printing Log, kept during the mission.}
 \label{fig:3dprintinglog}
 \Description{The chart has several columns that describe important information about each print job, such as the name of the design, the version number, the purpose of the print, the astronaut requester, the materials and colors used, the status of the printing process, and whether the goal was achieved. There is also a column with links to pictures of the prints, both successful and unsuccessful. This chart helps the crew organize their work, learn from their mistakes, and make improvements for future missions}
\end{figure*}

\subsection{Study Setup}

This experiential action research was conducted as part of a 14-day simulated lunar mission in ICE (isolated, confined and extreme) conditions. The mission's daily schedule was pre-defined and consisted of research activities and extravehicular activities (EVAs), simulating moon walks in spacesuits. The 5-person crew of English-speaking volunteers from Poland and Germany, aged between 20-45, was tasked with performing EVAs every second day. The EVA missions simulated the transition of humans into non-habitat space in which, like real astronauts, the crew wore suits designed to make movement restricted and difficult to simulate equipment used for spacewalks on the lunar surface (see Fig. \ref{fig:teaser}). The crew also had to maintain and improve their equipment, based on their current needs. Consequently, they were given access to two Prusa 3D printers and complete freedom to use them. To design the prints the team used Autodesk Fusion 360. For translating the model they used PrusaSlicer. This practice of equipping isolated labs with 3D printers is established, as 3D printing is looked at as one of the alternative ways of providing necessary equipment to both analog and real astronauts \cite{en15249322}, In this study we were interested in the experience and best practices for becoming more familiar and efficient with 3D printing when exposed both to it, and to clear areas of need resulting from unique circumstances. We facilitated this process with a 3D printing workshop, which was conducted by the crew member most experienced with 3D printing, as part of the regular sessions that were included in the day's schedule. It was a short series of meetings in which crew members met in the Atrium of the outpost and participated first in a lecture on modeling and 3D printing and then in brainstorming sessions to elicit their current needs \cite{gwamuri:hal-02113460}. Hence, the crew organically, in a self-determined way, came up with ideas for prints, based on their unique situation and needs. In the study we wanted to evaluate how to expedite the learning process, encourage community members to take part in it, and how to limit 3D printing waste in a learning environment, where mistakes are expected \cite{SCHELLY2015226}. We also wanted to explore ways in which to make 3D printing more sustainable and hazard-free. Hence, after describing the 3D printing cases we provide guidelines based on the experience gathered in the study and the observations of its participants.

\subsection{Documentation of each 3D print project}
The crew documented printed designs to systematize the work (see Fig. \ref{fig:3dprintinglog}), motivate the results, and give a clean, simple report so that later they could reflect upon it, and draw lessons from the failed ones, and the entire process. They kept track of the following data:
\begin{itemize}
    \item Design name - the working name of the print project, to recognize what it is and its purpose.
    \item Version number - to monitor the iterations that were made during the mission.
    \item Goal of the requested print - which is a description of what the print was made for and its purpose.
    \item Astronaut number(s)o of print requesters.
    \item Material(s) and color(s) - to - if necessary - repeat the design with stronger materials such as ABS or PETG.
    \item Status and note - specifying whether the printing process itself was a success, given that this is a technology that often fails during operation if not properly prepared, and noting the reasons it failed.
    \item Goal success - describing whether the previously set goal was achieved.
    \item Image file of the print - with links to the images of the prints, both successful and unsuccessful.
\end{itemize}

\section{Results and Discussion}

\subsection{Item 1: EVA BORP Suit Collar and Glove Rings} \label{borp}
\begin{figure*}[ht!]
   \begin{minipage}{1\textwidth}
     \centering
     \includegraphics[width=1\linewidth]{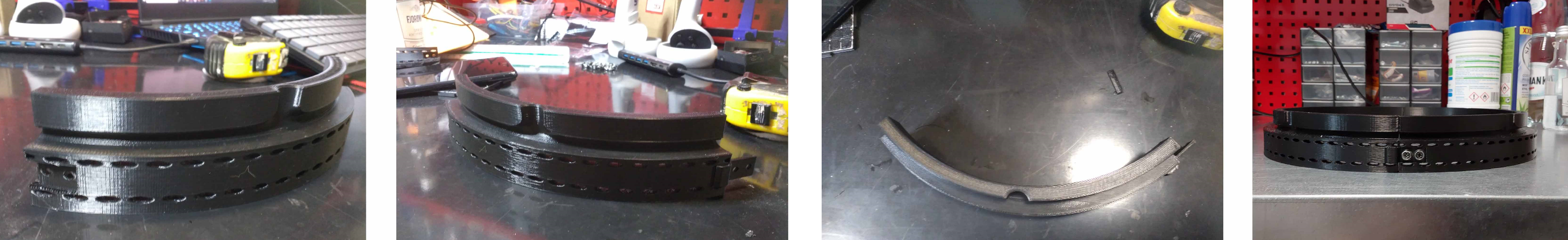}
     \caption{Right side of improved collar/Left side/View from Top/Connected collar}
     \Description{3D printed collar that is made from ABS material. It is black in color and is designed to be assembled from four parts. The collar was created to hold the helmet in place on an astronaut's suit during space missions. The four parts of the collar were put together to make a strong and durable piece that was almost the same size as the original but not exactly the same.}
     \label{borppic}
   \end{minipage}\hfill
\end{figure*}

The first thing the crew tackled was improving the design of the BORP Suit, a suit for analog space missions designed to maintain as much immersion as possible, designed in 2021 by the Sensoria 4 mission. The crew reported problems with the suit during the mission as the first EVA destroyed the collar designed to hold the helmet in place. Following the incident, all crew members analyzed the problem, especially the locking mechanism, and began repairs while looking for a structurally superior alternative. (see Fig. \ref{borppic})

Due to the lack of proper documentation on the scale of the FDM printed collar, they were unable to properly replicate the collar with stronger materials or to modify it. Despite attempts to "calculate" the scale and print it in the correct proportions, the crew was forced to repair the old flange using a soldering iron and an overabundance of materials. Eventually, the crew managed to produce an improved version of the flange from ABS that was close in size to the original. Nevertheless, it further deviated from the original scale \cite{https://doi.org/10.1002/admt.202000028}. 

Another problem was that the crew was diverse in terms of physique, height and weight which caused some complications during the preparation for EVA procedures. The rings connecting the BORP suit to the gloves were too small for more than half of the crew to squeeze their hands through. The Mission Engineer took a particular model of the safety ring from the previous crew, measured the dimensions, and printed larger rings for this purpose. The design itself also used a stronger material, ABS. As a result, more crew members were able to experience the full immersion of the suit BORP suit.

 \subsection{Item 2: EVA Suit: WUK-67} \label{wuk}
\begin{figure*}[htb]
   \begin{minipage}{1\textwidth}
     \centering
     \includegraphics[width=1\linewidth]{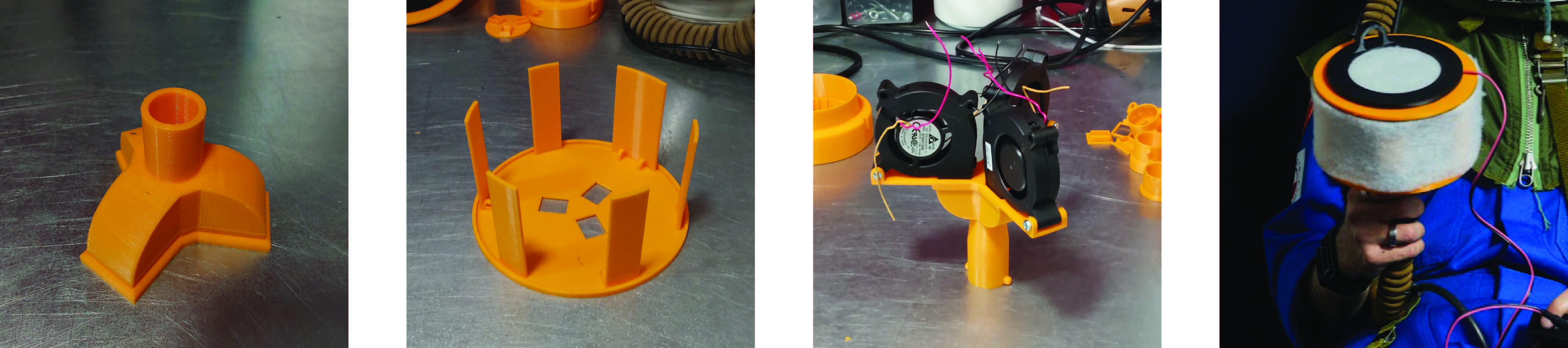}
     \caption{From left to right: redirector for air, case for fans/fans in the redirector and the completed Air Filter}
      \Description{Shown in the image are various components of a filter design that is intended to purify air extracted during simulated EVA (Extravehicular Activity) procedures from a confined space. The components include a prototype redirector for the air with the required inputs, a structural base for the filters and fans, the fans that are attached to the device, and the filter assembly itself that is linked to the WUK-67 air supply system.}
      \label{Redicrector}
   \end{minipage}\hfill
\end{figure*}

The second type of EVA suit was the Puma, which consisted of an incomplete WUK-67 pressure suit along with a collar, a Puma flight helmet, and a work suit. In addition to the structural problems, there was also the problem of oxygen delivery in the Puma suit as there was no system to force the air inside the helmet. The crew, after the first two EVAs began designing a component that would attach to the tube through which oxygen was originally delivered, so that the wearer would not suffer from hypoxemia while performing EVA activities. They began by prototyping direct-type fan inputs and an adapter to connect the component to the helmet. After several tests of the fans and the filter, a housing was designed and printed for the entire system (see Fig. \ref{Redicrector}) The crew, after mounting the system to the backpack responsible for the Support System, was able to efficiently use the new solution to deliver filtered air to the helmet.

 \subsection{Item 3: BioLab Plant Experiments} \label{biolab}
\begin{figure}[htb]
    \centering
    \includegraphics[width=\linewidth]{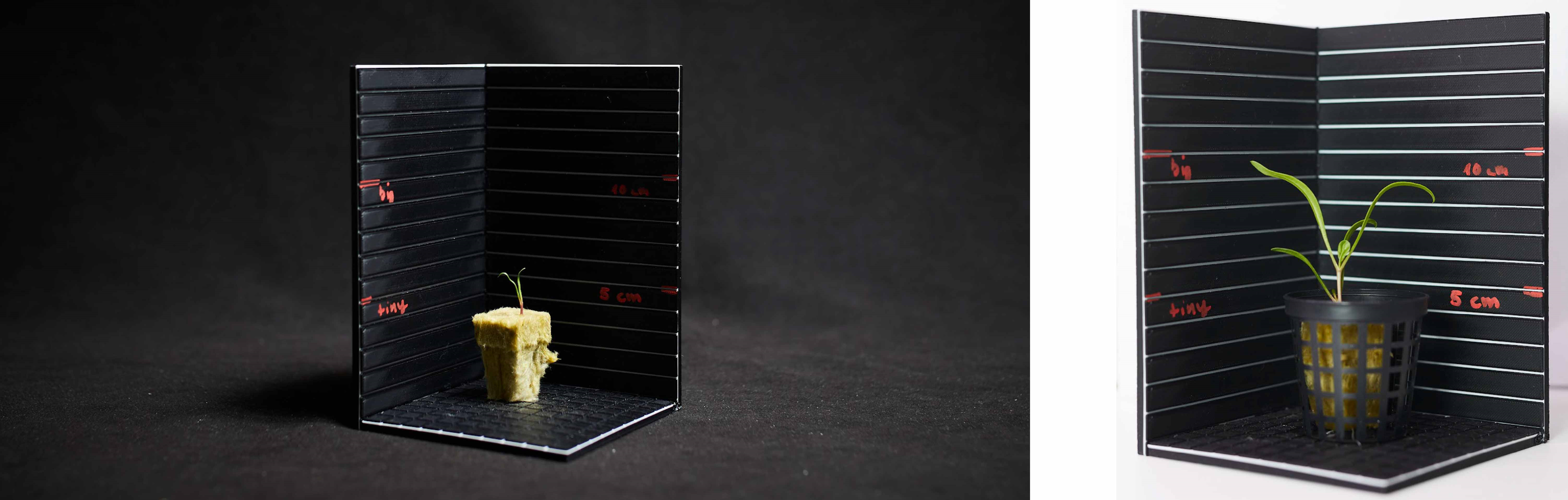}
    \caption{Photo booth at the beginning of the mission/Photo booth with one of the plants}
    \label{fig:booth}
     \Description{The picture depicts a photobooth comprising of three parts, which has gaps marked at intervals of 1 centimeter. The photobooth allows for manual marking of the planted bean, along with a leaf, at the start of the Alpha-XR mission. The same photobooth is also shown after two weeks of the mission, with the bean appearing to be significantly improved compared to its initial state.}
\end{figure}

Meanwhile, a BioLab was also available in which assigned crew members were engaged in comparative studies about the effects of water quality and nutrient composition on plant growth. It contained 4 tanks of water and 4 separate groups of beans, which were photographed daily as part of documenting the progress the plants were making under specific conditions. To improve the process a small photo booth was conceptualized to photograph, on daily basis, the growth progress of the plants reared in the BioLab. Finally, a small black photo booth with white accents was designed to indicate 1 cm distance between the lines (see Fig. \ref{fig:booth}). In the final days of the mission most of the plants have grown significantly, outgrowing the photo booth.

 \subsection{Item 4: Needs During the Simulated EVA Missions and the Design of the Universal Holding System } \label{holder}

For communications during EVAs, the crew used a backup shortwave radio system, especially in the airlock, which had very thick walls. To communicate during the main part of the mission, including EVAs on the simulated lunar surface, they set up a small local mumble server running on Wi-Fi. They received communications through tablets with which each crew member was equipped. From these earphones were fed, where the cable was routed under the suit and the collar up to the ears. At the beginning of the mission, only one crewman, out of the three who carried out the EVA procedure, was equipped with a special hold, so that the so-called "Mission Control", or HabCom, had access to the live video feed from the tablet and could make decisions efficiently. The rest of the tablets were attached to the suits with Velcro, which caused them to fall off frequently due to the lack of firmness of the suit. To this end, Mission Engineer co-designed a universal system through which the suit could be functionally split with elements predefined by the requirements of the EVA procedure. 

\begin{figure}[htb]
 \centering
\includegraphics[width=.4\textwidth]{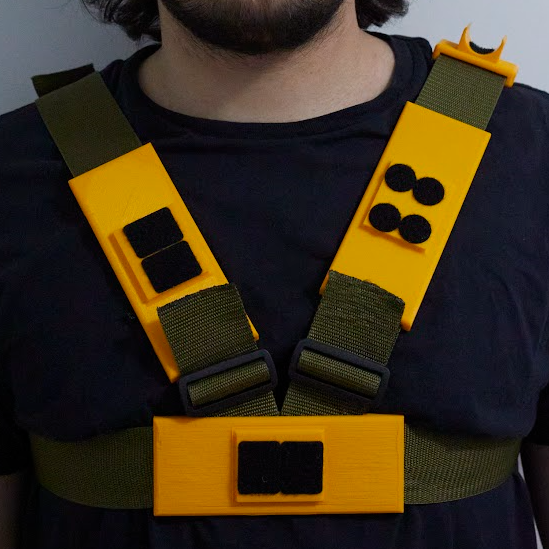}
 \caption{Universal Holding System}
 \label{fig:holdingsystem}
  \Description{The holding system is made up of adjustable straps and 3D printed parts that can be attached using Velcro. It is a versatile system that can be used to securely hold various equipment in place during EVA procedures.}
\end{figure}

The premise of the harness was that you could put anything in there that would be needed for an EVA procedure. The most important things were the tablets so that you could quickly pull them out, check the tablet, the contents, and put the tablet away, so that it would continue to attach smoothly to the body no matter what conditions the analog astronaut was in. To this end, a simple tablet holder was designed and printed into which the equipment could be inserted in such a way that the holder would not let go of the tablet until force was exerted on it. Along with the holder, an adapter was designed to which the holder could be attached. All the equipment was efficiently held on to the Velcro in such a way that dust and sand did not easily get into the Velcro and decrease its effectiveness.

\begin{figure*}[htb]
   \begin{minipage}{1\textwidth}
     \centering
     \includegraphics[width=1\linewidth]{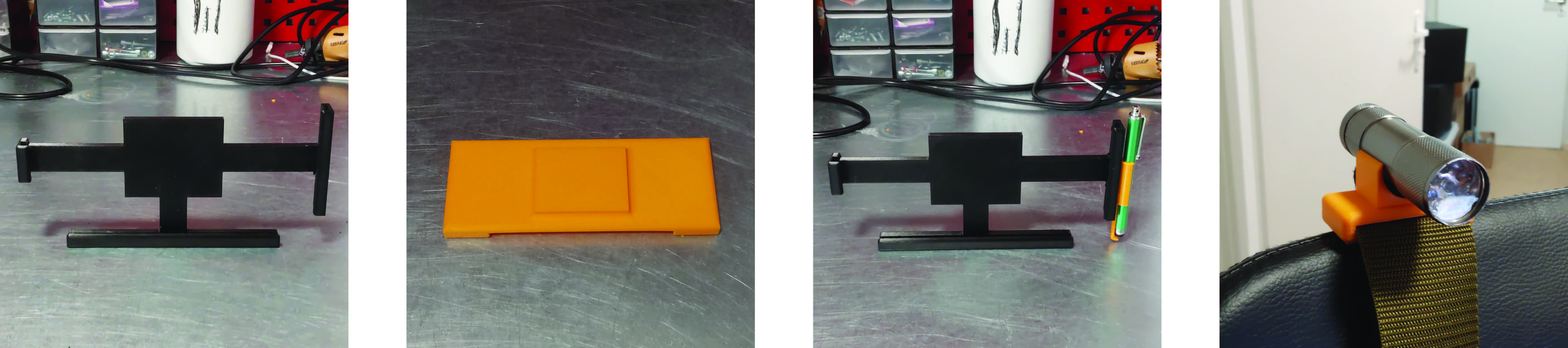}
     \caption{Tablet holder version 3/Attachment system/Holder with stylus/Flashlight}
      \Description{The tablet holder is a simple design that securely holds the tablet in place until force is applied. The improved design features shortened elements to allow for access to cable inputs and a stylus holder. User were also able to use flashlight.}
      \label{Tabletholder}
   \end{minipage}\hfill
\end{figure*}

During testing, after making the holder, it turned out that the design blocks the possibility of charging the tablet without removing it; moreover, it blocks a key element during the EVA sequence, i.e. the inputs for headphones. In the improved model, the elements holding the tablet were shortened, which allowed access to the cable inputs and saved printing time and material. In addition to the grip, a piece that holds a stylus to the tablet has also been designed. The presence of this small pen is important, because it is the only thing that gives you the ability to operate the tablet. The gloves of the suits in the habitat are so thick that the tablet screen simply does not respond to them - making the use of a stylus mandatory. Additionally, after the first night's EVAs a small holding device was designed to fit into the harness - which half of the EVA crew was already equipped with. It was based on the shape of the flashlights available in the habitat, and thanks to a small amount of Velcro, it was able to efficiently hold the flashlight in one place to light up the environment, freeing one hand when needed.

 \subsection{Item 5: EVA Suit: The DOC Outfit} \label{doc}
The last of the three EVA outfits was designed as the lightest, as the crew mate wearing it was in the support role for analog astronauts with heavier, more resistant, but movement-restricting, costumes. It was the only suit whose helmet did not fully protect against the dust in the EVA area, which simulated small fine dust called regolith of the lunar surface. It also lacked an air circulation system, causing condensation and decreased visibility. To this end, the crew tried to design an airflow solution using a windmill they found in the workshop section of the habitat. They designed an adapter to attach the windmill to the helmet. Combined with a power source hidden in the backpack responsible for the symbolic "life support system," the windmill was able to route contaminated air out of the now-closed helmet. Due to limited time in the habitat, the project was documented and this phase concluded with a precariously working prototype, with the goal of completing the project in future habitat missions. \cite{00003246-200410000-00016}

 \subsection{Item 6: The Bollywood Pin} \label{bolly}
It was a tradition to choose where to stay on the moon, which would define the light conditions for EVA procedures during the mission \footnote{The Moon is a natural satellite of the Earth and its cycle of days and nights is determined by its rotation around its axis and orbiting around the Earth. As we know, one full orbit of the Moon around its own axis takes about 29.5 Earth days (known as a synodic month), which means that it takes the same amount of time as one cycle of the Moon's phase (from new moon to new moon). In contrast, one orbit of the Moon around the Earth lasts about 27.3 Earth days (known as an anomalous month). These motions cause the observed side of the Moon to be variable as it orbits the Earth. As a result, we alternate seeing successive phases of the Moon, such as new moon, first quarter, full, third quarter. Consequently, for an observer on Earth, one side of the Moon is constantly facing the Earth, called the "side always facing the Earth" (the so-called attracting side), while the other, the turning side, is not visible from Earth. Therefore, we speak of an ongoing night and day on the moon.}. Consequently, due to the chosen location, more than half of the EVA procedures took place in the dark. It was also a tradition for the habitat to create a so-called Map Pin to mark the location of the mission. The crew designed a 3D-printed pin showing a man in a strange dancing position. This was to commemorate the daily routine of starting workout sessions with a "Bollywood Dance" used as a warm-up. This was important for the crew to facilitate the sense of building a community to show willingness to integrate and perform. Such displays of group spirit are important, especially in ICE conditions, that are prone to challenge the well-being of people exposed to them. \cite{https://doi.org/10.1111/jpi.12826}

\section{Discussion and Lessons Learned}

\subsection{Pollution and the Use of Sensors}
Pollution was the most significant challenge when it comes to the printers' operation. However, in a small, isolated space it is not limited to toxic gases, but also temperature and sound. FDM printing materials such as ABS when heated, begin to release toxic gases at the expense of better durability. These can cause nausea and headaches. Another problem is a rise in temperature as in order to produce models using FDM technology, the printer nozzle needs to heat up to 200+ degrees Celsius, while the large surface worktable needs to be more than 60 degrees Celsius. These sources of heat can change overall conditions in the area, affecting for example plant life. For these reasons good ventilation, including air filtration and cooling systems, was necessary. Despite the printers being locked off in an engineering workshop, which then had to be taken out of operation, smells associated with 3D printing were present in the habitat, as it was needed to check on the printers' operation as they worked. Another issue was the sound pollution, as working printers could be heard in the engineering lab, which had to be used for other tasks as well. A dashboard with data from air-pollution, temperature, humidity and sound sensors in a visible place next to the entrance to the workshop, together with a print-facing camera feed may have helped to limit the urge to check on the prints and may have kept the users more aware of the hazards involved. \cite{PINHEIRO20211}

\subsection{Print Longevity, Filaments, Storage and Recycling}
On-demand manufacturing and flexibility are among the key benefits of 3D printing technology -- prints can replace broken parts, but can also serve as temporary solutions to encountered problems or can be used to prototype completely new tools. Yet, this ease of prototyping and building tools to solve a one-time specialized problem creates a stack of failed, test or no-longer-in-use prints. In the simulated lunar habitat this stack was in the form of two large disorganized boxes with different unidentified prints, printed using diverse materials. This was not only wasteful, but also confusing as it lacked a knowledge-transfer element and left the current 3D-printing lab users to guess the original purpose of these tools and the reasons why they ended up in the discard box. For this reason, in such places it is necessary to have a filament shredder and extruder -- a setup that creates new filament out of 3D printing waste. To enable this, however, it is necessary to keep track of prints, including the materials used. This calls for efficient storage solutions, not only for fresh filament, which also ought to be described and stored neatly, but also for the prints to be recycled as these may affect the quality of recycled filament. Such recycling solution could aid in making the isolated community more self-sufficient, as some filament resupplies could be avoided.

\subsection{Co-Designing 3D Solutions with Requesters}

The need to 3D print solutions during the mission emerged organically from a few sources: (1) repairing failed equipment (See Section \ref{borp})
(2) improving quality of life for scheduled tasks, that is EVA missions (See Sections \ref{wuk}, \ref{borp}, \ref{doc})
(3) improving and supporting experiments (See Section \ref{biolab})
(4) facilitating the sense of community and creativity (See Section \ref{bolly}). One key observation is that it was not possible to tell at the beginning what specifically the crew members may need. The awareness of needs comes from both the experience of living in the specific ICE conditions, and the knowledge that these needs can be met with 3D printing. To facilitate this awareness we organized workshops to familiarize the crew with 3D printing technology. They were also helpful in discussing current shortages and areas where the procedures, tools and living conditions were lacking. This was possible, as the assigned Mission Engineer had some previous experience in 3D design. Despite this, the crew did not avoid some pitfalls they later reflected upon. For prototyping, when scale was not clear, the crew could have changed the design, to print just the element that was necessary to evaluate the scale, and not the whole print - which is really an afterthought, given the shortage of materials and time necessary to print the full design. Another related mistake was using the stronger materials before it was clear that the design is final -- the temptation to do this comes from printing times being long, however, as the stronger materials, such as ABS, may need to be kept for other purposes. Outsourcing some of the designs to external experts in such circumstances may have also been beneficial, again, to limit the number of unusable prints. Yet, while the isolated community members, such as crews in similar habitats would have little problem searching for information and cooperating on designs with Mission Support teams or External Contractors using the network, in the future, the crews of space missions, e.g. on Mars, may have a problem searching for information or collaborating on designs real-time, having a synchronization delay of up to 48 minutes. We have discovered a real value in the hands-on approach of co-designing necessary parts by the crew themselves as many ideas appeared or were fine tuned during the digital design phase, facilitated by the Mission Engineer. Another interesting effect was that the more time passed, the more members of the crew became involved in co-designing solutions, as they became familiar with the technology and the possibilities it offers (this can be seen in the Excerpt from the 3D printing log, Fig. \ref{fig:3dprintinglog}).
For teams thinking of using 3D printers in similar conditions we recommend in-depth training for one team member, who can not only maintain the printers, but also facilitate design and hold educational workshops for other community members, to evaluate, gather and help them meet their needs, both immediate ones related to repairs, and quality of life improvements.

\section{Conclusions}

Despite the decades-long history of 3D printing, this technology is still not used to its full potential. Our experiential study revealed three types of barriers to entry: setting up and maintaining the 3D printing equipment, design skill and familiarity with software and the awareness of what can be achieved with it, while minding its limitations. Despite these, as we explored, it also holds much promise for small isolated communities, which are facing resupply challenges; thus, ought to rely on in-community resources to repair and bring in new equipment. In-community experience and know-how are reliable sources of 3D print ideas, that may improve quality of life of community members. Their unique know-how is something external support services may have trouble tapping into. Another unique aspect is connected to the physical setup of the 3D printing lab that needs to not only limit different kinds of pollution, that is air, temperature and sound, but also make users aware of possible health hazards, including equipment handling. Therefore, co-design of 3D prints and facilitating awareness of what 3D printers can do for such communities is a promising area of study, that can bring with itself new applications of 3D print technology, especially in small, specialized communities.


\bibliographystyle{ACM-Reference-Format}
\bibliography{bibliography}










\end{document}